\documentclass[prl,reprint,flushbottom,showpacs]{revtex4-1}
\usepackage{epsfig}
\usepackage{amsmath}
\usepackage{amssymb}
\usepackage{amsthm} 
\usepackage{bm}
\usepackage{color}

\newcommand{\ket}[1]{\left|#1\right>}

\newcommand{\nn}{\nonumber\\}

\newcommand{\bea}{\begin{eqnarray}}
\newcommand{\ea}{\end{eqnarray}}
\newcommand{\eea}{\end{eqnarray}}
\newcommand{\ord}{{\cal O}}
\newcommand{\sumint}[1]

\begin{document}

\title{Emergence of a new pair-coherent phase in many-body quenches of repulsive bosons} 
\author{Uwe R. Fischer, Kang-Soo Lee, and Bo Xiong}

\affiliation{Seoul National University,   Department of Physics and Astronomy \\  Center for Theoretical Physics, 
151-747 Seoul, Korea}

\begin{abstract}
We investigate the dynamical mode population statistics and associated 
first- and second-order coherence of an interacting bosonic 
two-mode model when the pair-exchange coupling is quenched from negative to positive
values. 
It is shown that for moderately rapid second-order transitions, a new pair-coherent 
phase emerges on the positive coupling side in an excited state, 
which is not fragmented as the ground-state single-particle density matrix would 
prescribe it to be. 
\end{abstract}


\maketitle

The fundamental question whether an interacting system of bosons 
is a single or fragmented condensate \cite{Penrose,Leggett,Nozieres,Mueller} 
has conventionally been answered in favor of the single Bose-Einstein
condensate. This conclusion is gained from highly symmetric (e.g. rotationally 
invariant) situations, for example 
in antiferromagnetic spinor condensates 
\cite{Ho} or rotating gases \cite{UedaLeggett}.   
The 
fragmented condensate states occurring for these symmetrical Hamiltonians 
are extremely fragile against decay into a coherent state (a single condensate) by small perturbations coupling the single-particle modes \cite{Jackson}, breaking, e.g., rotational symmetry. 
As a consequence, the fragmented states are also extremely sensitive to small time-dependent perturbations like number fluctuations and generally excitations above the fragmented condensate ground state. Therefore, the instability 
of fragmentation towards a single condensate was traditionally assumed to be generic. 
However, a recent result within the two-mode approximation for scalar 
bosons with general interaction couplings (that is when one moves away from points of high symmetry), has shown that a continous 
variety of ground state fragmentation can be obtained in a single trap \cite{Bader}. 
This type of fragmentation is, importantly, robust to perturbations coupling on the 
single-particle level.

In the following, we investigate whether the robustness of continous
fragmentation also persists against rapid changes on the 
{\em dynamical many-body} 
level, i.e. when interaction couplings rapidly change.
Due to their many-body origin which is relying
on the (relative) values of the couplings, 
fragmented states might then be expected to be more fragile.
Here, we show that the sensitivity of fragmentation strongly depends on the range 
covered by a given coupling sweep. 
For a second-order dynamical quantum phase transition from the 
coherent to the fragmented phase \cite{Dziarmaga}, the creation of single-trap fragmentation from a coherent state is possible only for very slow sweeps 
of the coupling of bosonic pair exchange essentially  
determining the character of the ground state (single versus fragmented condensate \cite{Bader}). For moderately rapid exponential 
sweeps, 
the final degree of fragmentation is suppressed. The single-particle coherence measure $g_1$ (when averaged over time) vanishes after the sweep as well, while an analogously defined pair-coherence measure $g_2$ is macroscopically large. 
We therefore obtain a new pair-coherent phase which shows no single-particle 
fragmentation, in contrast to the ground state. 
The emergence of this phase 
is due to crossing the singular point of vanishing pair-exchange coupling. On the other hand, 
for sweeps entirely on the positive exchange-coupling side, no excited-state 
suppression is obtained even for large sweep rates, and the 
degree of fragmentation remains close to its ground-state 
value. We stress that these phenomena cannot occur in the fragmented phase 
of a conventional double-well with no pair exchange between sites included \cite{Streltsov}.

Let us consider the following two-mode Hamiltonian describing 
the dynamics of particles in an arbitrary trap and for a 
two-body interaction  
\cite{Bader} 
\bea
\begin{split}
\hat H =  
\sum_{i=0,1} \epsilon_i  \hat n_i
+\frac{A_1}2 \hat n_0 (\hat n_0-1)
+\frac{A_2}2 \hat n_1 (\hat n_1-1)  \\
+\frac{A_3 }{2}\left(\hat a_0^\dagger\hat a^\dagger_0
\hat a_1\hat a_1+ {\rm h.c.}\right) +\frac{A_4}2 \hat n_0 \hat n_1
. \label{H} 
\end{split} 
\ea
The Hamiltonian may be understood to result from going one step beyond the 
familiar semiclassical Gross-Pitaevskii theory \cite{Leggett}, i.e., by inserting the field operator decomposition  $\hat \Psi = \hat a_0 \Psi_0 + \hat a_1 \Psi_1 +\cdots$
into the full second-quantized Hamiltonian and truncating the expansion after
the first two modes. We thus neglect $\ord(1)$ fluctuations on top of these  
(now generally still quantum) modes, which are assumed to be the only modes macroscopically 
populated with a finite fraction of the number of particles, $N_i=\ord{(N)}$. 
The modes and two-body interactions are inhomogeneous and anisotropic in an independent way, leading to a set of interaction couplings $\{A_i\}$ with essentially 
arbitrary relative magnitudes and signs.
We assume that the modes $\Psi_i$ have been determined, e.g., by the multiorbital mean-field method delineated in \cite{StreltsovLong}, and integrated out. 

The sign of the pair-exchange coupling constant 
$A_3$ basically decides upon the classes of many-body ground states 
which can be obtained from the two-mode model \eqref{H} \cite{Bader}.
In contrast to optical lattices with pair-hopping \cite{Trotzky,Eckholt,Liang}, 
here both the couplings $A_3$ and $A_4$ have 
generally equal importance as $A_1$ and $A_2$. 
The Hamiltonian \eqref{H} then describes, e.g., single-trap fragmentation 
in harmonically trapped quasi-1D and quasi-2D gases in the weakly confining directions \cite{Fischer}.

In what follows, we assume for simplicity that the 
single-particle states are degenerate, $\epsilon_0 = \epsilon_1$, 
so that we can omit the first term in \eqref{H}; this does not change our results 
in any essential way.
We employ the following ansatz for the many-body wave function 
\bea
\ket \Psi = \sum_{l=0}^N \psi_l(t) \frac{(a_0^\dagger)^{N-l}}{\sqrt{(N-l)!}} \frac{(a_1^\dagger)^l}{\sqrt{l!}} 
\ket 0 
\equiv \sum_{l=0}^N \psi_l(t) \ket l ,
\label{ansatz} 
\ea 
which represents a linear superposition of Fock states 
$\ket{N-l,l} \equiv\ket l $. 
We are interested in the time-dependence of the occupation amplitudes $\psi_l(t)$, which comprise the information on the quantum  many-body dynamics. 
At any instant, $|\psi_l(t)|^2$ gives the probability that $l$ particles are in 
the second mode and $N-l$ in the first. 
Both the amplitude and phase of the $\psi_l$ will influence the temporal behavior 
of the first- and second-order coherence functions $g_1$ and $g_2$, to be described below.

The degree of fragmentation of an interacting many-body system, when maximally two field operator modes are macroscopically occupied, can  be defined in an invariant manner from the difference of the (macroscopic) eigenvalues of the single-particle density matrix \cite{Penrose,Bader} 
\bea
{\cal F} = 1-\sqrt{1-\frac4{N^2} \left(N_0 N_1  
- |\langle \hat a_0^\dagger \hat a_1 \rangle|^2\right)}\,.
\label{F}
\ea 
Here, $N_i  = \left\langle \hat n_i \right\rangle$ are the diagonal elements of the  single-particle density matrix, while the first order coherence 
$g_1 = \frac12 \langle \hat a_0^\dagger \hat a_1  + \hat a_1^\dagger \hat a_0 \rangle$ contains the real part of the off-diagonal correlator. For a coherent 
state with relative phase $\Delta\vartheta$, $g_1 = \sqrt{N_0 N_1} \cos[\Delta\vartheta]$, and $\cal F$ vanishes.

We assume that the interaction couplings $A_i$ 
are large enough for the system to attain a fragmented state \cite{Fischer},  
and that the couplings fulfill the condition $A_1+A_2+2|A_3|-A_4 >0\,\, \forall\,\, t$ 
\cite{Bader}. 
To set the stage for the dynamical case and introduce some notions, 
we first discuss the stationary ground state solution for the $\psi_l$ amplitudes.
Due to structure of the Hamiltonian, even and odd $l$ sectors decouple from each other
for $A_3\neq 0$ (cf.\,Eq.\,\eqref{transition} below),
and one gets a solution of the matrix equations for 
$\psi_l$ with a state 
for even and odd $l$ separately,
$	
|\phi \rangle = \sum_{\textrm{even}\,l}{\phi_l |l \rangle},\,\, 
	|\Phi \rangle = \sum_{\textrm{odd}\,l}{\Phi_l |l \rangle}. 
$
These two states are degenerate 
in the large $N$ limit. 
Consequently, any complex choice 
of $a$ and $b$ in the superposition 
$\ket{\Psi}= a\ket{\phi}+b\ket{\Phi},\, |a|^2+|b|^2=1$ 
leads to a ground state at the same energy, 
with vanishing first-order coherence $g_1$ and generally 
nonvanishing fragmentation 
$\cal F$ on the $A_3>0$ side.
There results, for $A_3>0$,  
a purely imaginary (or vanishing) 
correlator  $\langle \hat{a}_0^{\dag} \hat{a}_1 \rangle 
	= \sum_{l} \omega_l \psi_l^* \psi_{l+1}$, 
\begin{equation}
\begin{split}
	\langle \hat{a}_0^{\dag} \hat{a}_1 \rangle 
	&= \sum_{\textrm{odd}~l}{ 
	\omega_{l-1} a^* b \, \phi_{l-1}^* \Phi_{l} 
	+ \omega_{l} a b^* \phi_{l+1} \Phi_{l}^* } \\ \label{a0a1}
	&\simeq 2 i |ab| \sin \theta
	\sum_{\textrm{odd}~l}{\omega_{l} |\Phi_l|^2 }, \\
\end{split}
\end{equation}
with $\omega_l \equiv \sqrt{ (l+1)(N-l) }$ and we defined the relative phase between even 
and odd sectors $\theta = \arg(b) - \arg(a)$. 
The second line is valid in the continuum approximation of slowly varying occupation amplitudes, $|\Phi_l| \simeq |\phi_{l\pm 1}|$, 
and $\omega_l \simeq \omega_{l\pm 1}$.
Hence $g_1$, the real part of the above correlator, 
vanishes in the continuum approximation. 
 
The degree of fragmentation \eqref{F}, as a function of 
$|a|$, $|b|$, and the relative phase $\theta$ between $a$ and $b$ 
(which were assumed to be real in \cite{Bader}) then evaluates to
\bea
{\cal F} = 
1- \frac 2N \sqrt{\left[|ab| N \sin\theta \left(1-\frac{\sigma^2 +
2{\mathfrak S}^2}{N^2}\right)\right]^2\!+{\mathfrak S}^2} . \label{Ftheta}
\ea
We used the Gaussian ground state distribution of 
$|\Phi_l| \simeq |\phi_{l\pm 1}| \simeq (\sqrt{\pi /4} \sigma )^{-1/2} \exp [ -(j-\mathfrak{S})/(2 \sigma^2 ) ]$ representing a solution in the continuum limit [4, 8]. We defined $j = l - \frac{N}{2}$, $\mathfrak{S} = \frac{N}{2} - N_0 
= N_1-\frac N2$ as the shift from a
maximally fragmented state and the distribution width
$\sigma = \sqrt{N} R^{1/4}$ with $R(\{A_i\})=\frac{|A_3|}{A_1+A_2+2|A_3|-A_4}$
(when ${\mathfrak S} \ll N/2$).
For $\theta=\pi/2 +\delta$ ($|\delta|\ll 1$)
and $|a|=|b|=1/\sqrt2$, the degree of fragmentation is suppressed to 
${\cal F} = \sqrt{R}/N + \delta^2/2$, while being maximal for a given set $\{A_i\}$ 
at $\theta=0\,\,({\rm mod}\,\,\pi)$.  Adding a small 
(e.g. $\Omega=\ord{(A_1/N)}$)  
Josephson type perturbation 
\bea
\hat H_{\rm J} = - \frac{\Omega} 2 \left( \hat a_0^\dagger\hat a_1 + {\rm h.c.}\right)
\label{HJ}
\ea 
on the $A_3<0$ side, the system is driven to a coherent state with $a=b$ and 
${\cal F}=0$. On the $A_3>0$ side, the ground-state fragmentation is insensitive to $\Omega$, the correction to $\cal F$ being of order $\Omega^2/(NA_3)^2$, provided the singular region around $A_3=0$ is excluded. 

We now proceed to the dynamical case, using $A_1\equiv 1 $ as unit of energy in the following. 
We assume an exponential sweep of the form 
$A_3(t) = (A_i-A_f) \exp[-\alpha t] + A_f$ with $A_i = A_3(t=0)$
and $A_f=A_3(t\rightarrow\infty)$, and that all other couplings 
remain essentially constant during the sweep $\partial_t A_3 \gg 
\partial_t A_1,\partial_t A_2,\partial_t A_4$ \cite{DD}.
The matrix equations resulting from \eqref{H}, \eqref{ansatz} and \eqref{HJ} 
($\hbar=1$)
\begin{eqnarray} \label{transition}
      i \partial_t \psi_{l}
      & = &  \frac{A_{3}(t)}2 \left[ d_{l}\psi_{l+2} + d_{l-2}\psi_{l-2}\right]   +c_{l}\psi_{l}\nn
      & & -\frac{\Omega}2 \left[\omega_{l} \psi_{l+1} - \omega_{l-1} \psi_{l-1} \right]  
   \end{eqnarray}
are solved numerically,  where the coefficients $c_l = \frac12A_1(N-l)(N-l-1)
+\frac12A_2 l(l-1)+\frac12A_4 (N-l)l$ and  $d_l = \sqrt{(l+2)(l+1)(N-l-1)(N-l)}$.

\begin{center}
\begin{figure}[t]
\centering
\hspace*{-1em}
\includegraphics[width=.48\textwidth]{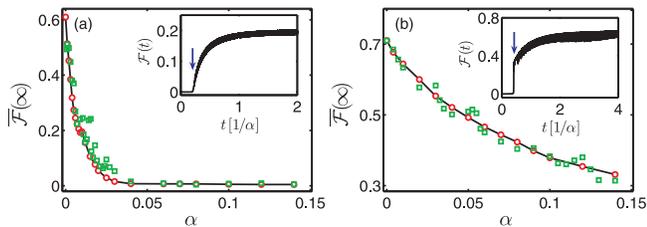}
\caption{\label{Fragalpha} Average asymptotic value of degree of fragmentation
versus sweep rate $\alpha$ across $A_3=0$. Red circles are from numerical solutions of 
\eqref{transition} for $\Omega=0$ and solid lines a guide for the eye.
For a second-order transition 
(a) $A_1 = A_4 =1, A_2 = 0.5, A_i=A_3(t=0)=-0.2, A_f =A_3(\infty)=0.4, 
N=100$, ground state final value of ${\cal F}_{\rm gs} = 0.6$, and for (weakly) first order in 
(b) they are $A_2=0.6$,  $N=200$, ${\cal F}_{\rm gs} = 0.71$, $\Delta {\mathcal F} =1/3$, others identical.
The green squares 
are obtained by adding a perturbation of the form \eqref{HJ} with $\Omega =1$.  
The inset shows ${\cal F} (t)$ 
for $\alpha=0.01$;  
the blue arrow indicates where $A_3=0$.
Note the different vertical axis origin for (a) and (b).
}
\end{figure}
\end{center}

The Josephson parameter $\Omega$ is only dynamically effective in a small window around $A_3=0$ according to the ratio $\Omega/N A_3(t)$ [cf. the ground state considerations above], and its effect is correspondingly small even for slow sweeps across the 
phase transition at $A_3=0$; we illustrate this insensitivity 
by showing additional $\Omega =1$ data in Fig.\,\ref{Fragalpha} (green squares). 
While for second-order transitions, some small resonance peaks appear for slow sweeps [cf.\,Fig.\,\ref{Fragalpha}\,(a)], at sufficiently large values of $\alpha$,  
the identical complete suppression of fragmentation occurs.
We have verified that taking $\Omega=1$ also does not qualitatively 
alter the other $\Omega=0$ results to follow.

During the sweep,  the relative phase between the even and odd sectors, 
$\theta \equiv \frac2{N} \sum_{k=0}^{N/2}\{\arg (\psi_{2k+1}) - \arg(\psi_{2k})\},\,\forall \,\,|\psi_l|^2>0$, becomes time dependent. 
We will see that this has a crucial influence
on the final degree of fragmentation
$\cal F$ for sweeps from the coherent $A_3<0$ to the fragmented $A_3>0$ side.  
In the latter quantum phase transition, the 
singular point $A_3=0$ is crossed. At this point of vanishing 
pair-exchange coupling, the matrix problem in \eqref{transition} becomes 
diagonal in the $\ket l$ Fock-space, and thus is easily solved.  
One Fock state $\ket l$ is obtained, with a fragmentation jump 
$\Delta {\cal F} = 1- |\frac{A_1-A_2}{A_1+A_2-A_4}|$. 
The transition is therefore generally of first order, with a discontinous change of 
fragmentation  $\Delta{\cal F}$ at $A_3=0$. 
We explore here both the second- and first-order cases of the dynamical 
quantum phase transition \cite{Ralf} when crossing the singular 
$\psi_l$-distribution point at $A_3=0$. 

\vspace*{-2.25em}

\begin{center}
\begin{figure}[t]
\centering
\hspace*{-0.5em}
\includegraphics[width=.42\textwidth]{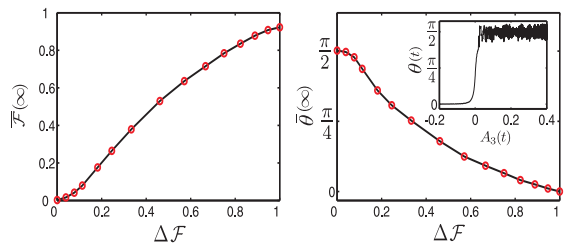}
\vspace*{0.5em}
\caption{\label{DeltaF} Left: 
The final degree of fragmentation for $\alpha=0.1$ as a function 
of the jump at $A_3=0$;  $N=200$ and the variation of $\Delta {\cal F}$
is achieved by varying $A_2$ from $\frac12$ to 1. Right: The average phase difference
at late times versus $\Delta {\mathcal F}$. The inset shows that in the second-order 
case $\Delta F=0$, the average $\theta$ approaches $\pi/2$ at late times. Other parameters 
are identical with those of Fig.\,\ref{Fragalpha}, $\Omega=0$.}
\end{figure}
\end{center}

We observe a strong dependence on there being a ground-state fragmentation jump at $A_3=0$, i.e. on the transition being second ($\Delta {\cal F}=0$) or first order 
($\Delta {\cal F}\neq 0$). The final average degree of fragmentation is defined as 
$\bar{\cal F} (\infty) \equiv \frac1{t_2 - t_1} \int_{t_1}^{t_2} {\cal F}(t') dt'$, 
where $t_2\gg t_1 $ are two late times well after the phase transition point is crossed. 
For $\alpha \gg 1/N$, the final degree of fragmentation $\bar {\cal F} (\infty)$
quickly tends to zero in the second-order case (Fig.\ref{Fragalpha}\,(a)).
On the other hand, we obtain that $\bar{\cal F}(\infty)$ decays much less rapidly with sweep rate $\alpha$ in the first-order case, cf. Fig.\ref{Fragalpha}\,(b).  We show the dependence of $\bar {\cal F} (\infty)$ on $\Delta {\cal F}$ in the left plot of Fig.\,\ref{DeltaF}.
For the second-order case (a) in Fig.\ref{Fragalpha}, the phase difference 
between the even and odd $l$ sectors is rapidly driven towards an average value 
$\bar\theta\simeq \pi/2$ after crossing $A_3=0$ for intermediate values of $\alpha$ 
(cf.\,Fig.\,\ref{DeltaF}).
The phase difference of 90 degrees between even and odd $l$ sectors
explains the strong suppression of fragmentation in this regime of 
$\alpha\ll 1$, while for first-order transitions, $\bar\theta$ well 
after the transition is significantly less than $\pi/2$. The 
reason for this intimate relation between $\bar {\mathcal F} (\infty)$ and $\bar \theta (\infty)$ is that the argument leading to the ground-state result 
\eqref{a0a1} is still approximately valid when the condition
$\arg(\psi_{l+1}) - \arg(\psi_{l-1})\simeq \pi\,({\rm mod}\,2\pi)$ is fulfilled 
for most $l$ values, a property which we have verified.
\vspace*{-1em} 


\vspace*{-2em}

\begin{center}
\begin{figure}[b]
\centering
\includegraphics[width=.32\textwidth]{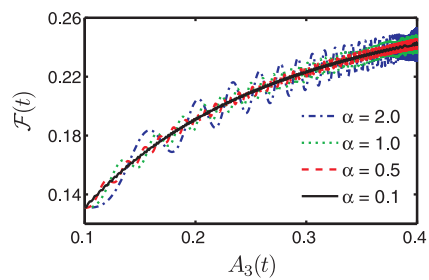}
\caption{\label{FragA3>0} 
Degree of fragmentation
for a sweep entirely on the $A_3>0$ side for increasing sweep rate, where 
$A_1=A_4=1, A_2=0.5, N=200, \Omega=0$. 
Even for large sweep rates of order $A_1$, ${\cal F} (t)$ remains
close to its ground state value, as well as $\theta$ close its initial value 
$\pi/4$;  $|a|=|b|=1/\sqrt2$ is constant.
}
\end{figure}
\end{center} 
\vspace*{-2em} 

The above behavior for the phase-transition sweep needs to be contrasted 
with the case $A_3$ strictly positive
during the whole sweep, imposing that $A_3 \gg \ord{(1/N)}$. 
The average degree of fragmentation is not suppressed, even  
when $\alpha\gg 1 $ (and thus 
larger than the interaction couplings $A_i$). We illustrate this 
in Fig.\,\ref{FragA3>0}; cf. the fragmentation 
suppression observed for the quench case displayed in Fig.\,\ref{Fragalpha}. 
The insensitivity of fragmentation therefore persists for dynamical changes
of $A_3$, provided the singular region around $A_3=0$ is not traversed. 

\vspace*{-2em}

\begin{center}
\begin{figure}[tb]
\centering
\includegraphics[width=.48\textwidth]{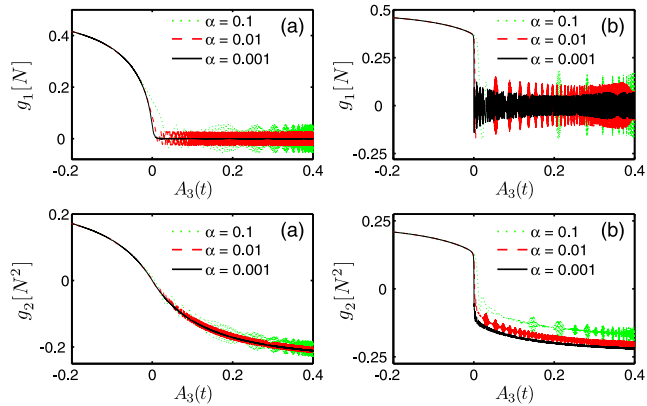}
\hspace*{-0.5em}
\caption{\label{g1g2} Dynamical behavior of first- and second-order 
pair coherence $g_1(t)$ (top) and $g_2(t)$ (bottom) for three different sweep rates $\alpha$. In the first-order case (b), the pair coherence decreases 
for larger $\alpha$ while it remains on average identical in the second-order case (a). Parameter sets are identical to those of Fig.\,\ref{Fragalpha}, with $\Omega=0$.} 
\end{figure}
\end{center} 

We now discuss the pair-exchange coherence, measured by the expectation value   
\bea
g_2 \equiv \frac12 \left\langle \hat a_0^\dagger \hat a_0^\dagger 
\hat a_1 \hat a_1 + \hat a_1^\dagger \hat a_1^\dagger \hat a_0 \hat a_0
\right\rangle . 
\ea  
We observe, first of all, that the quantity $g_2$ can be macroscopic, i.e., $g_2\sim\ord(N^2)$, when the many-body state is neither coherent nor fragmented.
In the ground state, it can be shown by direct calculation that for either sign of $A_3$, there is pair coherence except right at $A_3=0$ (where $g_2$ crosses zero), and that $g_2$ is independent of the weights $a$ and $b$. In particular, we have $g_2 \simeq -N^2/4 +\Omega^2/(8N^2A_3^2)+\sigma^2/2+{\mathfrak S}^2$ for $A_3>0$. 

The dynamical behavior of $g_2$ across the sweep, along with that of its single-particle counterpart $g_1$ is shown in Fig.\,\ref{g1g2}. Whereas the degree of fragmentation is  
suppressed (see also Fig.\,\ref{Fragalpha}), and the time-averaged value of first-order coherence is zero, an essentially stationary pair condensate emerges, cf.\,\,the bottom row of Fig.\,\ref{g1g2}.
For a (weakly) first-order transition (in the case shown, $\Delta {\mathcal F} = 1/3$), fluctuations
become stronger  and there is an increasing  
suppression of pair-coherence for larger $\alpha$.

While on the single-particle coherent side, $g_1\neq 0$, positive pair-exchange coherence is trivially achieved due to the existence of first-order coherence, on the $A_3>0$ side 
pair-coherence {\em without} single-particle fragmentation is the manifestation of a new pair-correlated phase emerging after the sweep. 
We emphasize that the pair coherence does not result from attraction between bosons,
as all $A_i$ are chosen positive on the $A_3>0$ side \cite{Pashitskii}.


We have shown 
that for moderatedly rapid variations of the pair-exchange coupling from negative to positive values in a second-order quantum phase transition, a new pair-coherent phase is created. In contrast to ground-state expectations,
the resulting many-body state is not a single-particle fragmented state, with the microscopic origin that the average phase difference between even and odd sectors of the many-body amplitudes approaches $\pi/2$ after the quench. 
Rapid temporal changes of the interaction couplings of a many-body system can thus result in an emergent quantum phase with coherence 
properties strikingly different from those of the adiabatic ground state.

This research was supported by the 
Brain Korea
BK21 program and the 
NRF of Korea, grant No. 2010-0013103. 


\begin{thebibliography}{9999}
\bibitem{DD} The condition that temporal variations of $A_3$ are more rapid than those of the other 
couplings can, for example, be realized by nonlocal and anisotropic
interactions like the dipole-dipole interaction, cf. the extended footnote of Ref.\,\cite{Bader}, or by suitable imaginary mode functions.
\end{thebibliography}
\end{document}